\documentclass[usenatbib,usegraphicx,useAMS]{mn2e} 

\usepackage{txfonts}
\usepackage{xspace}
\usepackage{listings}
\usepackage{color}
\usepackage{caption}
\usepackage{times}
\usepackage{url}
\usepackage{lineno}

\newcommand{\kepler}{\textit{Kepler}\xspace}
\newcommand{\corot}{\textit{CoRoT}\xspace}
\newcommand{\gimenez}{Gim\'enez\xspace}
\newcommand{\pytransit}{\textit{PyTransit}\xspace}

\title[PyTransit]{PyTransit\\ Fast and Easy Exoplanet Transit Modelling in Python}
\author[H. Parviainen]{Hannu Parviainen\\University of Oxford, Sub-Department of Astrophysics, Department of Physics, Oxford OX1 3RH}

\begin{document}

\maketitle
\begin{abstract}
We present a fast and user friendly exoplanet transit light curve modelling package \pytransit, implementing 
optimised versions of the \cite{Gimenez2006} and \cite{Mandel2002} transit models.  The package offers an 
object-oriented Python interface to access the two models implemented natively in Fortran with OpenMP 
parallelisation. A partial OpenCL version of the quadratic Mandel-Agol model is also included for GPU-accelerated 
computations. The aim of \pytransit is to facilitate the analysis of photometric time series of exoplanet 
transits consisting of hundreds of thousands of datapoints, and of multi-passband transit light curves from 
spectrophotometric observations, as a part of a researcher’s programming toolkit for building complex, 
problem-specific, analyses.
\end{abstract}

\section{Introduction}
\label{sec:introduction}
The rapid increase in computational power during the last decades has allowed for the adoption of increasingly robust 
statistical methods in the analysis of astrophysical data. Specially, combining a fully Bayesian approach to inference 
with Markov Chain Monte Carlo (MCMC) sampling for the posterior estimation has allowed for improved characterisation of 
model parameter uncertainties in parameter estimation problems, while the adoption of Bayesian model selection has given 
us the tools to robustly judge between many competing hypotheses aiming to explain our observational data. The increased 
flexibility and robustness come with a price. While the methods allow us to work with complex models with a high number 
of dimensions (free  parameters), increasing dimensionality quickly increases the number of likelihood evaluations 
required for the analysis. Further, while the computers keep getting faster, also the size of the observational datasets 
(and the reserachers' ambitions) keep growing thanks to the advancements in instrumentation and observation techniques.

In the analysis of photometric times series of exoplanet transits (transit light curves), the size and complexity of the 
observational datasets has increased due to the introduction of space-based telescopes \corot and \kepler, observing 
potentially hundreds of individual transits for a single transiting planet;\footnote{And, while the \kepler long-cadence 
(LC) mode produces a relatively small number of exposures per transit, the modelling of long cadence data requires 
model supersampling to account for the extended integration time \citep{Kipping2010a}.} due to 
introduction of lucky-imaging techniques allowing for very high time resolution observations; and due to the maturing of 
spectrophotometry as a transit observation method. 

The space-based telescopes and lucky-imaging cameras produce light curves with tens to hundreds of thousands exposures, 
while the ground-based spectrophotometric observations of individual transits yield a smaller number of exposures, but 
with an additional dimension (number of passbands extracted from the observed spectra) to our time series. Both the 
stellar limb darkening and the planetary radius vary as a function of the wavelength coverage of the passband, and in 
typical cases the dimensionality of the parameter space increases by 3-5 free parameters per passband (radius ratio, at 
least two limb darkening parameters, assuming we are not overly reliant on the theoretical limb darkening coefficients, 
and 1-2 parameters to model the baseline flux). This increase in dimensionality leads to a significant increase in the 
number of likelihood evaluations needed to obtain a representative posterior sample using MCMC techniques, and, thus, an 
analysis of a single spectrophotometric transit can require equal amounts of computation resources as an analysis of a 
\kepler light curve covering hundreds of transits.

The transit shape model forms the core of the forward model in the transit light curve analysis. The model describes
the dependence of the observed flux as a function of stellar limb darkening, planet-star radius ratio, and the
planet-star distance (from centre to centre, expressed in stellar radii.) The field of transit light curve modelling is
largely dominated by two analytical approaches: a set of transit models for different stellar limb darkening
parametrisations by \citet{Mandel2002}, and a versatile series-expansion-based model presented by \citet{Gimenez2006}.
Both the \gimenez and Mandel-Agol models (henceforth G and MA models, respectively) allow the most computationally
intensive calculations to be factored out from the effects due to limb darkening, accelerating the calculation of 
multiple simultaneously observed passbands with different limb darkening coefficients significantly.

Here we present a Python package offering a straight-forward way to access the \citet{Gimenez2006} and 
\cite{Mandel2002} transit models.  The package has been used for transit light curve modelling in nine
peer-reviewed publications over four years, and can be considered production ready. The models are implemented in
Fortran~2003 (based on the original FORTRAN77 implementations by the respective authors) with OpenMP multithreading and
model-specific optimisations aimed to minimise the model evaluation cost. Both models can be computed exactly, or using
interpolation for improved speed, and a partial OpenCL implementation of the interpolated quadratic Mandel-Agol model is
included for GPU computing. The package includes the necessary utility routines to calculate circular and elliptic
orbits (using either Newton's method, iteration, or two series approximations), transit durations, eclipse centres,
etc., and offers a simple interface combining the orbit and the transit model computations that selects the most
appropriate orbit calculation routine based on the eccentricity. Examples and tutorials on using the code are included
in the package and online.\footnote{See the \texttt{notebooks} directory from \url{https://github.com/hpparvi/PyTransit}
for IPython notebook examples, and \url{https://github.com/hpparvi/exo_tutorials} for more in-depth tutorials on
exoplanet characterisation in general.}

While the MA and G models are the two most commonly used approaches for transit modelling, both more 
generic and specialised modelling tools exists. \citet{Abubekerov2013} have derived analytical expressions 
for the transit shape (and its derivatives, also derived by \citet{Pal2008} for the quadratic MA model) for a wider set 
of limb-darkening models than offered by \citeauthor{Mandel2002}, and offer an example implementation written in~C. The 
\nobreak{JKTEBOP} package by \citet{Southworth2008} offers a versatile numerical approach to transit modelling where 
both the host star and the planet are modelled as biaxial spheroids. This goes beyond basic transit shape model, 
allowing for the modelling of the reflection and ellipsoidal effects as a function of orbital phase. \citet{Barnes2009} 
has introduced a numerical approach for modelling transits over rapidly-rotating stars with significant 
gravity-darkening due to stellar oblateness. The transits over rapidly rotating stars can be use to probe for 
spin-orbit misalignment, since misaligned orbits will show asymmetric transits. Finally, \citet{Pal2012} have 
considered the problem of modelling mutual transits in multi-planet systems, something the MA or G models cannot be 
used for directly.

{\pytransit aims to offer a Pythonic access to the tools for one part of the planet characterisation problem: 
the modelling of the flux decrement due to an occulting planet as a function planet-star distance, planet-star radius 
ratio and stellar limb darkening. 
Several other transit modelling packages offer similar functionality with their own advantages and limitations. 
EXOFAST by \citet{Eastman2013} is an IDL library for transit and radial velocity modelling.\footnote{With a 
web-front-end at \url{http://astroutils.astronomy.ohio-state.edu/exofast/exofast.shtml}} The authors gain significant 
improvements in the evaluation speed of the quadratic MA model by swapping the original method for computing the 
elliptic integral of the third kind with a faster one. However, the use of IDL, while still relatively popular in 
astrophysics, limits the package's adoptability.\footnote{However, \citeauthor{Eastman2013} also offer Python and 
Fortran implementations of their faster MA model.
The Transit Analysis Package \citep[TAP,][]{Gazak2011} is an IDL transit modelling package with a graphical user 
interface. The package implements the wavelet based likelihood function by \citet{Carter2009} that accounts for 
correlated noise (of very specific statistical characteristics) in the photometry, improving the robustness of the 
parameter estimates.
Finally, the latest update of PlanetPack \citep{Baluev2014} has included transit modelling using the 
\citeauthor{Abubekerov2013} transit model and several correlated noise models. PlanetPack is a command-line program 
written in C++, and thus its use has the pros and cons of a an analysis approach based on a single monolithic program. 
However, this is alleviated by its open-source nature combined with its independence from proprietary packages and 
languages.
}

\pytransit advocates the toolkit-based approach where the analysis code is constructed using a set of tools best suited 
for the problem at hand. This is similar to EXOFAST (although the scope of the package is significantly narrower), and 
lower-level than what offered by TAP and PlanetPack. The approach offers significant flexibility what comes to working 
with different MCMC samplers, optimisers, implementing noise models, etc., but it also requires the end-user to have a 
slightly higher level of experience than what is required by the off-the-shelf analysis packages.

\section{The Transit Models}
\label{sec:transit_models}
\subsection{The \gimenez Model}
\label{sec:gimenez_model}
\subsubsection{Overview}

\citet{Gimenez2006,Gimenez2007} describe a versatile series-expansion based transit model developed originally for 
eclipsing binaries by \citet{Kopal1977}.  The normalised flux, $f$, is expressed as
\begin{equation}
 f(k,z) = 1 - \alpha(k,z), \label{eq:gimenez_flux}
\end{equation}
where $\alpha$ stands for the fractional loss of light due to the transiting planet, $k$ is the star-planet radius 
ratio, and $z$ is the projected star-planet distance divided by the stellar radius. The $\alpha$ functions are described 
as \begin{equation}
 \alpha(k,z) = \sum_{n=0}^N C_n \alpha_n(b,c), \label{eq:gimenez_alpha}
\end{equation}
where $C_n$ are factors that depend only on $n$ limb darkening coefficients, 
$b= k/(1+k)$, and $c=z/(1+k)$, and the $\alpha_n$ functions are expressed in terms of Jacobi polynomials, $G_n(p,q;x)$ as
\begin{eqnarray}
 \alpha_n(b,c) &= \frac{b^2 (1-c^2)^{\nu+1}}{\nu \Gamma(\nu+1)} \times \; \sum_{j=0}^{N_J} (-1)^j (2j+\nu+2) 
                  \frac{\Gamma(\nu+j+1)}{\Gamma(j+2)}  \nonumber \\
               &\times \; G_j(\nu+2, \nu+1; 1-b)^2 \times \; G_j(\nu+2, 1; c^2), \label{eq:gimenez}
\end{eqnarray}
where $\nu = (n+2)/2$, and $\Gamma$ is the gamma function.

The accuracy of the \gimenez model is determined by $N_J$, the number of Jacobi polynomials used in 
Eq.~\ref{eq:gimenez}.  Figure~\ref{fig:gimenez_error} shows the model's maximum and mean absolute deviations from the 
exact solution for $z=[0..1+k]$ as a function of $N_J$. The figure is only illustrative, since the exact deviations 
depends on the limb darkening and the spanned $z$-values (but the dependency on limb darkening is small relative to the 
dependency on $N_J$.) Relatively small $N_J$ ($\sim$40) can be found to be sufficient for modelling ground-based 
observations. Also, a small $N_J$ can be first used to first obtain a quick rough solution, which can then be refined by 
increasing $N_J$.

\begin{figure}
 \centering
 \includegraphics[width=\columnwidth]{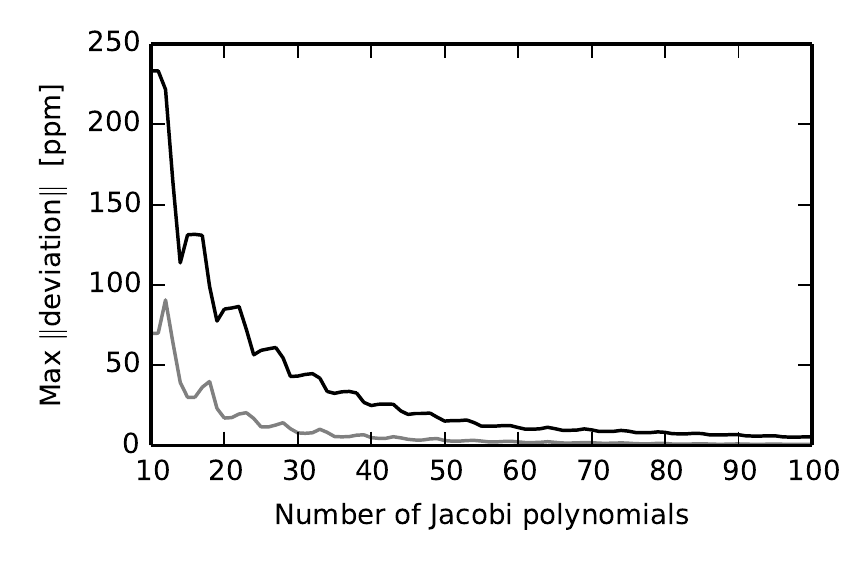}
 \caption{Maximum and mean (black and grey lines, respectively) absolute deviations for the \gimenez model with two limb 
darkening coefficients as a function of number of Jacobi polynomials used.} \label{fig:gimenez_error}
\end{figure}

\subsubsection{Limb darkening}
The \gimenez model produces transit light curves that follow a general limb darkening law,
\begin{equation} \label{eq:ld_general}
 \frac{I(\mu)}{I(1)} = 1 - \sum_{n=1}^N u_n (1-\mu^n), 
\end{equation}
where $N$ stands for the order of the model, $u_n$ is the n$^{th}$ limb darkening coefficient, $\mu = \cos \gamma$, and 
$\gamma$ is the foreshortening angle (the angle between the surface normal and the line of sight.) The general law 
equals to linear limb darkening for $N=1$,
\begin{equation}
 \frac{I(\mu)}{I(1)} = 1 - u(1-\mu),
\end{equation}
and to the quadratic limb darkening law for $N=2$,
\begin{equation}
 \frac{I(\mu)}{I(1)} = 1 - a(1-\mu) - b(1-\mu)^2,
\end{equation}
where the quadratic law coefficients ($a,b$) are transformed to the general law coefficients as $u_1 = a+2b$ and $u_2 = -b$.

\subsubsection{Optimisation: Simultaneous Multiple Passbands}
\label{sec:gimenez_multicolor}

The limb darkening enters into the \citeauthor{Gimenez2006} model as coefficients 
\begin{equation}
 C_0 = \frac{1-\sum_{n=1}^N u_n}{1-\sum_{n=1}^N \frac{n u_n}{n+2}},
\end{equation}
\begin{equation}
 C_n = \frac{u_n}{1-\sum_{n=1}^N \frac{n u_n}{n+2}},
\end{equation}
where $C_n$ do not depend on $k$ or $z$. The coefficients are cheap to compute, and by first calculating the $\alpha_n$ 
in Eq.~\ref{eq:gimenez_alpha}, the model can be evaluated with low computational cost for different sets of limb 
darkening coefficient vectors $\mathbf{u}_i$ as \begin{equation}
 \alpha_i(k,z) = \sum_{n=0}^N C_{n,i} \alpha_n(b,c). \label{eq:gimenez_alpha_multicolor}
\end{equation}
This is beneficial for the analysis of spectrophotometric data, since we do not need to evaluate the full model for each 
separate passband.

\subsubsection{Optimisation: Precomputing the Model Coefficients}
\label{sec:gimenez_optimisation}

A quick look at the Eq.~\ref{eq:gimenez} shows that the computationally most expensive parts of the equation do not 
depend on $k$ or $z$, but only on the depth of the expansion and the number of limb darkening coefficients. If we 
abbreviate \begin{equation}
 A_{n,j} = \frac{(-1)^j (2j+\nu+2)}{\nu \Gamma(\nu+1)} \frac{\Gamma(\nu+j+1)}{\Gamma(j+2)}, \label{eq:gimemez_a}
\end{equation}
Eq.~\ref{eq:gimenez} can be written as
\begin{equation}
 \alpha_n(b,c) = b^2 (1-c^2)^{\nu+1} \sum_{j=0}^\infty A_{n,j} G_j(\nu+2, \nu+1; 1-b)^2 \; G_j(\nu+2, 1; c^2),
\end{equation}
and $A_{n,j}$ can be precomputed into a two-dimensional lookup table given the number of limb darkening coefficients and 
the depth of the series expansion in the beginning of the computations. 

Next, the Jacobi polynomials $G_j$ can be calculated using recursion as
\begin{eqnarray}
 G_0(q,p;x) =& 1, \nonumber \\
 G_1(q,p;x) =& ( (2+q+p) x + q-p ) / 2, \\
 G_{i+1}(q,p;x) =& ( ( j_1 + j_2 x ) G_i + j_3 G_{i-1} ) / j_4, \nonumber
\end{eqnarray}
where the coefficients $j_n$ also depend only on the expansion depth and number of limb darkening coefficients, and can 
be precomputed into another two-dimensional lookup table.

After $A_{n,j}$ and $j_n$ have been precomputed, the evaluation of the Gim\'enez model for a given radius ratio, 
projected distance, and limb darkening coefficients requires only summations and multiplications, making the computation 
of the model for a large number of points fast.

\subsection{The Quadratic Mandel-Agol Model}
\label{sec:ma}
\subsubsection{Overview}
\label{sec:ma:overview}
\citet{Mandel2002} introduced a set of analytical transit light curve models for several different limb darkening laws, 
of which \pytransit implements the uniform and quadratic model. The flux, $f$, for a transit over a stellar disk with 
quadratic limb darkening is 
\begin{equation}
 f(k,z) = 1 - \frac{(1-c)\lambda_e(k,z) + c\lambda_d(k,z) + b\epsilon_d(k,z)}{1-a/3-b/6}, \label{eq:ma_model}
\end{equation}
where $k$ is the radius ratio, $z$ is the projected distance, $c=a+2b$, $a$ and $b$ are the quadratic limb darkening 
coefficients, and $\lambda_e$, $\lambda_d$, and $\eta_d$ are functions that depend on $k$ and $z$ as defined in 
\citet{Mandel2002}.

\subsubsection{Optimisation: Simultaneous Multiple Passbands}
\label{sec:ma:multicolor}
As with the \citeauthor{Gimenez2006} model, the effects from limb darkening are factored out from the most expensive 
computations (those of $\lambda_e$, $\lambda_d$, and $\eta_d$), which, again, allows for efficient computation of 
multiple simultaneous passbands with different limb darkening coefficients.

\subsubsection{Optimisation: Precomputing Interpolation Tables}
\label{sec:ma:interpolation}
The functions $\lambda_d$, $\lambda_e$ and $\eta_d$ in Eq.~\ref{eq:ma_model} can be precomputed into two-dimensional 
interpolation tables spanning $z=[0\,..\,1+k]$ and $k$ ranges based on the prior set on $k$. The model evaluation can 
now be done by first interpolating the values of the three functions, followed by the summations, multiplications, and 
one division (two of the divisions can be replaced with multiplications) in Eq.~\ref{eq:ma_model}.

The maximum absolute error (here defined as the deviation from the non-interpolated model) for the interpolated
MA model using the default values ($n_\mathrm{k}=128$ and $n_\mathrm{z}=256$) for $\Delta k=0.02$ (that is, we have an
uniform prior on the radius ratio from $k_0$ to $k_0+\Delta k$) is $\sim$4~ppm, and the average absolute
deviation over the whole transit is 0.05~ppm. Both of these values are well below the limits that can be achieved
\kepler or \corot. Decreasing $n_\mathrm{k}$ has a smaller effect on the introduced error than decreasing
$n_\mathrm{z}$, and even $n_\mathrm{k}=8$ for $\Delta k=0.02$ yields a maximum absolute error of $\sim$8~ppm.

\section{Implementation}
\label{sec:implementation}
\subsection{Overview}
\label{sec:implementation:overview}
The transit models are implemented in Fortran~2003 based on the original FORTRAN77 implementations by 
\citeauthor{Gimenez2006} and \citeauthor{Mandel2002}. The shared-memory parallelisation is carried out using OpenMP. The 
Python package offers easy-to-use Python classes wrapping the Fortran models with a common interface for both models. 
The Python interface offers automatic model supersampling given the number of subsamples and integration time, and model 
interpolation in $z$-space is implemented for the \gimenez model (the use of interpolation tables for the quadratic MA 
model makes any gains of $z$-space interpolation insignificant.)

\subsection{Model Supersampling}
\label{sec:implementation:supersampling}
A single point in a photometric time series corresponds to an integration of the flux over the exposure time. If the 
changes in the observed signal are small compared to the noise level over a single exposure, we can approximate this 
integration with a single value evaluated at the centre of the exposure. However, this approximation fails when the 
exposure time is long enough to integrate over modelled features---such as with \textit{Kepler}'s long time cadence mode 
\citep{Kipping2010a}---and also the model needs to be integrated over the exposure.

\pytransit offers transit model supersampling to account for long exposure times given the number of subsamples and the 
exposure time. The model is evaluated for $n$ evenly distributed subsamples inside each exposure, where $n$ is 
given by the user based on exposure time and transit duration, and each light curve point corresponds to the mean of 
the subsamples.

\subsection{\gimenez Model Interpolation in z-Space}
\label{sec:implementation:gimenez_interpolation}
While the optimisations described in Sec.~\ref{sec:gimenez_optimisation} makes the Gim\'enez model computationally 
efficient, simple model interpolation in $z$-space can offer a notable speedup with light curves containing hundreds of 
thousands of points. 

The code implements an alternative interpolated mode for the evaluation of the \gimenez model, where the transit model 
is first evaluated for $n$ points for $z<1-k$ and $m$ points for $1-k < z < k+1$, and the light curve points are 
interpolated from these tabulated values. The maximum absolute error (again, the difference between the
interpolated and non-interpolated models) for the default grid size is $\sim$25~ppm, and the average error is
$\sim$2~ppm.

\subsection{Partial OpenCL Acceleration of the MA model}
\label{sec:implementation:opencl}
The package implements an OpenCL version of the interpolated quadratic Mandel-Agol model, where the 
interpolation of  $\lambda_e$, $\lambda_d$, and $\eta_d$ is offloaded to the GPU. Given the computational simplicity of 
bilinear interpolation used, the overheads from memory transfer to and from the GPU dominate the evaluation time, making 
the model evaluation in GPU significantly slower than in CPU for small light curves. For basic usage, the OpenCL 
accelerated model is faster than the multithreaded Fortran implementation for light curves with $> 2\times10^5$ points. 
However, significant speedups can be reached with smaller light curves if both the $z$ and likelihood calculations are 
also offloaded to the GPU to minimise the memory transfer.

\subsection{Applicability to Transmission Spectroscopy}
\label{sec:implementation:applicability}

Both the \gimenez and Mandel-Agol models allow for efficient evaluation for multiple simultaneously observed passbands, 
where the differences in the transit shape (and observed depth) reflect the differences in the stellar limb darkening in 
each passband. However, this does nothing to include the effects from the variations in the effective radius ratio $k$, 
the parameter of interest when carrying out transmission spectroscopy. 

The effects on the transit depth by varying $k$ are included in \pytransit by assuming that the relative changes in the 
effective radius ratio are a small fraction of its average value. Now, the changes in $k$ affect only the transit depth, 
and the changes in other observables (the duration and shape of the transit) are below observation limits. Thus, given a 
set of $n$ radius ratios, $k_{i=1..n}$, the model is evaluated using the average radius ratio, $\hat{k}$, and then 
multiplied for each passband by a correction factor $k_i^2/\hat{k}^2$. 

\section{Performance}
\label{sec:performance}

We benchmark the model performance using two setups: 
\begin{enumerate}
 \item Intel Linux desktop (64 bit Ubuntu Precise), Intel i7-3770 (4 cores at 3.4 GHz), GFortran~4.6.3, optimisation 
    flags \texttt{-Ofast -march=native}. 
 \item AMD Linux desktop (64 bit Ubuntu Trusty), AMD FX-8350 (8 cores at 2.8 GHz), 
    GFortran~4.9, optimisation flags \texttt{-Ofast -march=native}.
\end{enumerate}
With the exception of threading, the performance scales equivalently for both setups (the absolute performance is also 
comparable when considering the difference in the clock rates), and we show the results only for the first setup.

Figure~\ref{fig:basic_scalability} shows the absolute model evaluation times as a function of the number of light curve 
points for the directly evaluated and interpolated transit models. The modelled light curves have 35\% of in-transit 
points and 65\% of out-of-transit points, corresponding to a typical observation setup. The interpolation for the 
\gimenez model is carried out using linear interpolation in $z$-space, while the interpolated Mandel-Agol model uses 
bilinear interpolation with the two-dimensional interpolation tables for $\lambda_d$, $\lambda_e$ and $\eta_d$. The 
quadratic Mandel-Agol model is significantly faster than the two-coefficient \gimenez model, but increasing the number 
of limb darkening coefficients does not affect the performance of the \gimenez model significantly. The OpenCL version 
of the interpolated Mandel-Agol model is slower than the Fortran version for small light curves due to memory transfer, 
but this can be alleviated by offloading also the rest of the computations to the GPU.

Figure~\ref{fig:evaluation_time_per_passband} shows the model evaluation times per passband relative to the evaluation 
time for a single passband. The speedup from calculating all passbands simultaneously when working with 
spectrophotometric data is obvious.

Figure~\ref{fig:threading} shows the evaluation time as a function of OpenMP threads for the setup with an Intel i7 
processor. The multithreading scaling is slightly different for the two setups: the optimal performance is obtained with 
three threads for the Intel setup and six threads for the AMD setup (not shown). 

\begin{figure}
 \centering
 \includegraphics[width=\columnwidth]{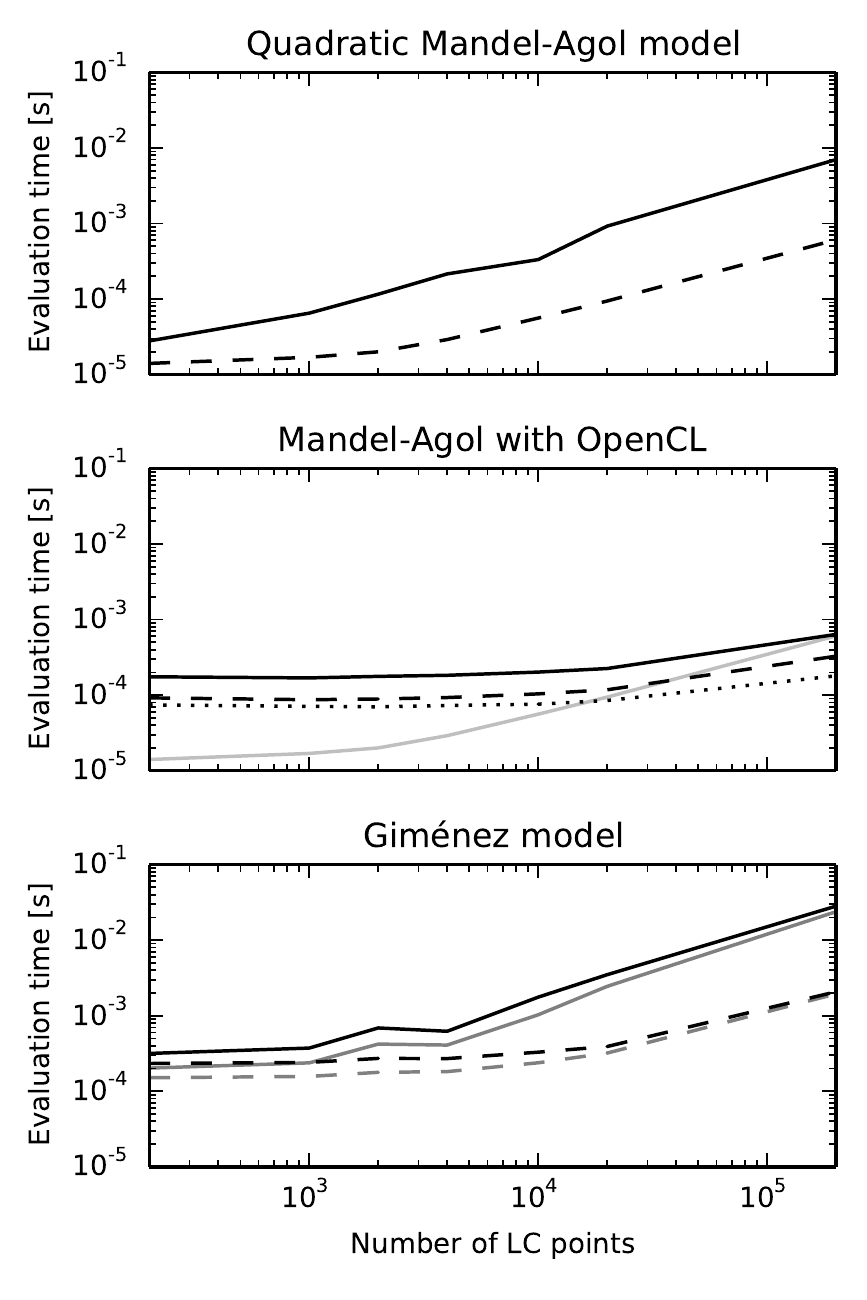}
 \caption{Model evaluation times for a single light curve with $n$ points. The topmost panel shows the directly 
evaluated MA model (solid line, using four threads), and the model interpolated in $\lambda_d$, $\lambda_e$ and $\eta_d$ 
(dashed line.) The middle panel shows the MA model with the interpolation carried out using GPU (coded in OpenCL): the 
black solid line corresponds to model evaluation with two memory transfers (first copying the $z$ array to the GPU, then 
reading the flux array from the GPU); the dashed line corresponds to model evaluation without transferring the flux from 
the GPU (the likelihood computation is also offloaded to the GPU); and the dotted line corresponds to model evaluation 
where also the z-array calculation is done in GPU. The solid grey line shows the CPU-interpolated MA model results for 
reference. The lowest panel shows the model evaluation times for the \gimenez model. The grey solid line shows the 
results for the directly evaluated quadratic model, the grey dashed line for the model interpolated in z-space, and the 
black lines correspond to the evaluation of four-parameter general limb darkening model.}
 \label{fig:basic_scalability}
\end{figure}

\begin{figure}
 \centering
 \includegraphics[width=\columnwidth]{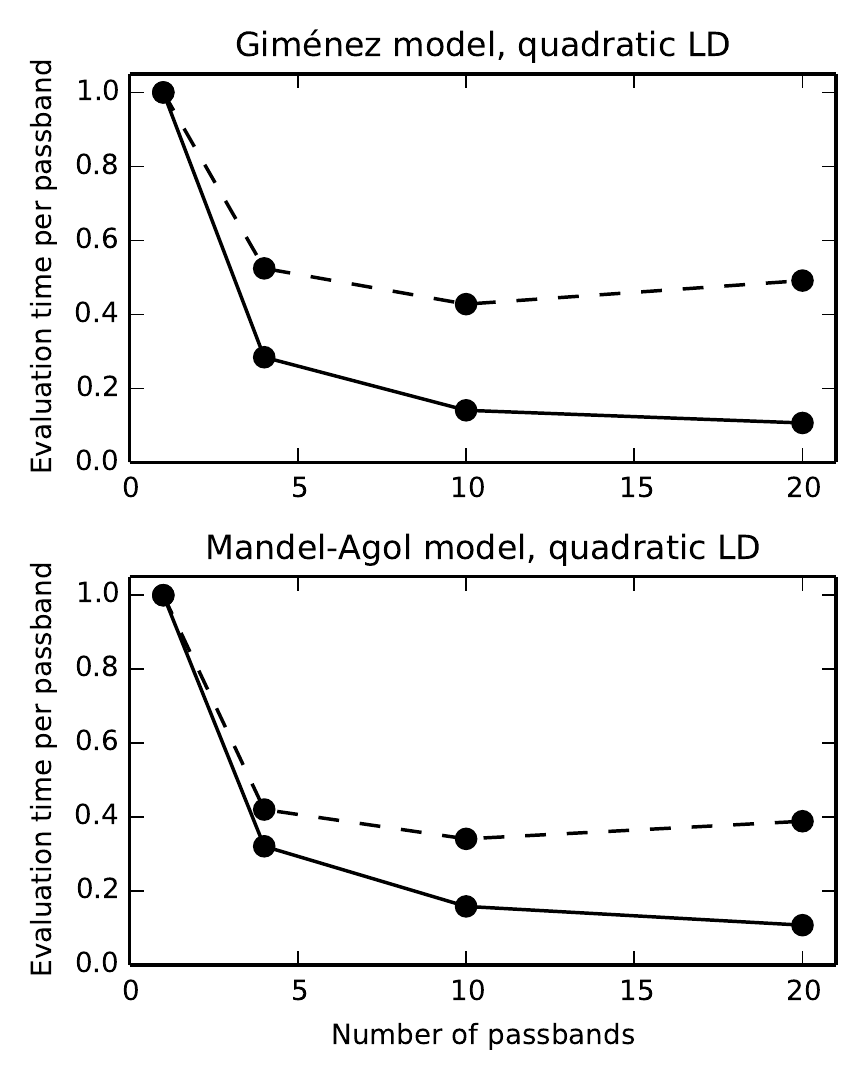}
 \caption{Model evaluation times as a function of simultaneous passbands relative to the evaluation time for a single 
passband. Continuous lines show the results for the direct models, and slashed lines for the interpolated models. Note 
that the gains for the interpolated models are smaller since the evaluation times for the parts not dependent on limb 
darkening are smaller.} \label{fig:evaluation_time_per_passband}
\end{figure}

\begin{figure}
 \centering
 \includegraphics[width=\columnwidth]{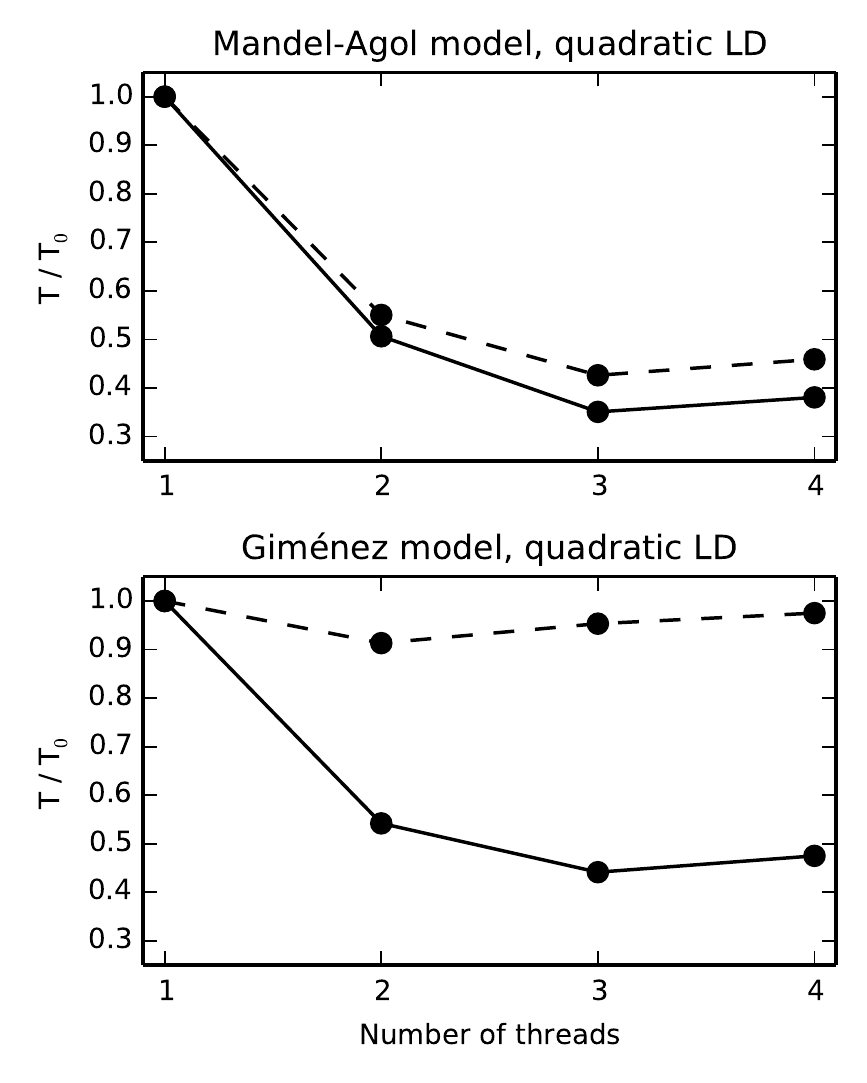}
 \caption{Model evaluation times for the Intel setup as a function of threads relative to the evaluation time of 
the respective model with one thread. Continuous lines show the results for the direct models, and slashed lines for 
the interpolated models.} \label{fig:threading}
\end{figure}

\section{Conclusions and Discussion}
\label{sec:conclusions}
We have described a Python package offering optimised versions of the \gimenez and Mandel-Agol transit models. The 
package is freely available from github
\begin{itemize}
 \item[] \url{https://github.com/hpparvi/PyTransit}
\end{itemize}
and under continuing development. The Fortran routines are also directly usable as Fortran modules. The
package comes with IPython notebook examples showing the use and features of the code. More in-depth tutorials
covering exoplanet characterisation from transit light curves can be found from
\begin{itemize}
 \item[] \url{https://github.com/hpparvi/exo_tutorials}
\end{itemize}
again implemented as IPython notebooks.

Interpolation can be used to speed up both of the models, and the performance gain is especially significant
for the MA model. The errors introduced by the interpolation are small, but systematic, and error evaluation is
recommended if extreme precision ($\sim 10^{-7}$ for the MA model, $\sim 10^{-5}$ for the G model) is required. The
maximum deviations for the interpolated MA mode occur at the end of ingress and the beginning of egress.

The two transit models have different advantages: the quadratic Mandel-Agol model is significantly faster than the 
two-coefficient \gimenez model, but the \gimenez model allows for higher-order limb darkening. Other Mandel-Agol models 
with higher-order limb darkening may be implemented in the future, but the flexibility of the \gimenez model makes this 
a low-priority task. 

The code has been developed and tested over several years, and has been used in 
\citet{Parviainen2014,Murgas2014,Gandolfi2014,Tingley2014a,Parviainen2012,Gandolfi2013,Murgas2012,Rouan2011a}; and
\citet{Tingley2011b}.

\section*{Acknowledgements}

The author warmly thanks H.~Deeg, J.A.~Belmonte and S.~Aigrain for their comments and helpful discussion, and the
anonymous referee for the constructive comments.
The development of the \pytransit package was supported by RoPACS, a Marie Curie Initial Training Network funded by the 
European Commission’s Seventh Framework Programme; by V\"ais\"al\"a Foundation through the Finnish Academy of Science 
and Letters; and by the Leverhulme Research Project grant RPG-2012-661.

\bibliographystyle{mn2e}
\bibliography{PyTransit}

\begin{thebibliography}{}

\bibitem[\protect\citeauthoryear{Abubekerov \& Gostev}{Abubekerov \&
  Gostev}{2013}]{Abubekerov2013}
Abubekerov M.~K.,  Gostev N.~Y.,  2013, MNRAS, 432, 2216

\bibitem[\protect\citeauthoryear{Baluev}{Baluev}{2014}]{Baluev2014}
Baluev R.~V.,  2014, arXiv:1410.1327

\bibitem[\protect\citeauthoryear{Barnes}{Barnes}{2009}]{Barnes2009}
Barnes J.,  2009, ApJ, 705, 683

\bibitem[\protect\citeauthoryear{Carter \& Winn}{Carter \&
  Winn}{2009}]{Carter2009}
Carter J.~a.,  Winn J.~N.,  2009, ApJ, 704, 51

\bibitem[\protect\citeauthoryear{Eastman, Gaudi \& Agol}{Eastman
  et~al.}{2013}]{Eastman2013}
Eastman J.,  Gaudi B.~S.,    Agol E.,  2013, PASP, 125, 83

\bibitem[\protect\citeauthoryear{Gandolfi, Parviainen, Deeg, Lanza, Fridlund,
  Moroni, Alonso, Augusteijn, Cabrera, Evans, Geier, Hatzes, Holczer, Hoyer,
  Kangas, Mazeh, Pagano, Tal-Or \& Tingley}{Gandolfi
  et~al.}{2015}]{Gandolfi2014}
Gandolfi D.,  Parviainen H.,  Deeg H.~J.,  Lanza a.~F.,  Fridlund M.,  Moroni
  P. G.~P.,  Alonso R.,  Augusteijn T.,  Cabrera J.,  Evans T.,  Geier S.,
  Hatzes a.~P.,  Holczer T.,  Hoyer S.,  Kangas T.,  Mazeh T.,  Pagano I.,
  Tal-Or L.,    Tingley B.,  2015, A\&A, 576, A11

\bibitem[\protect\citeauthoryear{Gandolfi, Parviainen, Fridlund, Hatzes, Deeg,
  Frasca, Lanza, {Prada Moroni}, Tognelli, McQuillan, Aigrain, Alonso, Antoci,
  Cabrera, Carone, Csizmadia, Djupvik, Guenther, Jessen-Hansen, Ofir \&
  Telting}{Gandolfi et~al.}{2013}]{Gandolfi2013}
Gandolfi D.,  Parviainen H.,  Fridlund M.,  Hatzes A.~P.,  Deeg H.~J.,  Frasca
  A.,  Lanza A.~F.,  {Prada Moroni} P.~G.,  Tognelli E.,  McQuillan A.,
  Aigrain S.,  Alonso R.,  Antoci V.,  Cabrera J.,  Carone L.,  Csizmadia S.,
  Djupvik A.~A.,  Guenther E.~W.,  Jessen-Hansen J.,  Ofir A.,    Telting J.,
  2013, A\&A, 557, A74

\bibitem[\protect\citeauthoryear{Gazak, Johnson, Tonry, Dragomir, Eastman, Mann
  \& Agol}{Gazak et~al.}{2012}]{Gazak2011}
Gazak J.~Z.,  Johnson J.~A.,  Tonry J.,  Dragomir D.,  Eastman J.,  Mann A.~W.,
     Agol E.,  2012, Adv. Astron., 2012, 1

\bibitem[\protect\citeauthoryear{Gim\'{e}nez}{Gim\'{e}nez}{2006}]{Gimenez2006}
Gim\'{e}nez A.,  2006, A\&A, 450, 1231

\bibitem[\protect\citeauthoryear{Gim\'{e}nez}{Gim\'{e}nez}{2007}]{Gimenez2007}
Gim\'{e}nez A.,  2007, A\&A, 474, 1049

\bibitem[\protect\citeauthoryear{Kipping}{Kipping}{2010}]{Kipping2010a}
Kipping D.~M.,  2010, MNRAS, 408, 1758

\bibitem[\protect\citeauthoryear{Kopal}{Kopal}{1977}]{Kopal1977}
Kopal Z.,  1977, Astrophys. Space Sci., 50, 225

\bibitem[\protect\citeauthoryear{Mandel \& Agol}{Mandel \&
  Agol}{2002}]{Mandel2002}
Mandel K.,  Agol E.,  2002, ApJ, 580, L171

\bibitem[\protect\citeauthoryear{Murgas, Pall\'{e}, Cabrera-Lavers, Col\'{o}n,
  Mart\'{\i}n \& Parviainen}{Murgas et~al.}{2012}]{Murgas2012}
Murgas F.,  Pall\'{e} E.,  Cabrera-Lavers A.,  Col\'{o}n K.~D.,  Mart\'{\i}n
  E.~L.,    Parviainen H.,  2012, A\&A, 544, A41

\bibitem[\protect\citeauthoryear{Murgas, Pall\'{e}, {Zapatero Osorio},
  Nortmann, Hoyer \& Cabrera-Lavers}{Murgas et~al.}{2014}]{Murgas2014}
Murgas F.,  Pall\'{e} E.,  {Zapatero Osorio} M.~R.,  Nortmann L.,  Hoyer S.,
  Cabrera-Lavers A.,  2014, A\&A, 563, A41

\bibitem[\protect\citeauthoryear{P\'{a}l}{P\'{a}l}{2008}]{Pal2008}
P\'{a}l A.,  2008, MNRAS, 390, 281

\bibitem[\protect\citeauthoryear{P\'{a}l}{P\'{a}l}{2012}]{Pal2012}
P\'{a}l A.,  2012, MNRAS, 420, 1630

\bibitem[\protect\citeauthoryear{Parviainen, Deeg \& Belmonte}{Parviainen
  et~al.}{2013}]{Parviainen2012}
Parviainen H.,  Deeg H.~J.,    Belmonte J.~A.,  2013, A\&A, 550, A67

\bibitem[\protect\citeauthoryear{Parviainen, Gandolfi \& Deleuil}{Parviainen
  et~al.}{2014}]{Parviainen2014}
Parviainen H.,  Gandolfi D.,    Deleuil M. e.~a.,  2014, A\&A, 562, A140

\bibitem[\protect\citeauthoryear{Rouan, Parviainen \& Moutou}{Rouan
  et~al.}{2012}]{Rouan2011a}
Rouan D.,  Parviainen H.,    Moutou C. e.~a.,  2012, A\&A, 537, A54

\bibitem[\protect\citeauthoryear{Southworth}{Southworth}{2008}]{Southworth2008}
Southworth J.,  2008, MNRAS, 386, 1644

\bibitem[\protect\citeauthoryear{Tingley, Palle, Parviainen, Deeg, {Zapatero
  Osorio}, Cabrera-Lavers, Belmonte, Rodriguez, Murgas \& Ribas}{Tingley
  et~al.}{2011}]{Tingley2011b}
Tingley B.,  Palle E.,  Parviainen H.,  Deeg H.~J.,  {Zapatero Osorio} M.~R.,
  Cabrera-Lavers A.,  Belmonte J.~a.,  Rodriguez P.~M.,  Murgas F.,    Ribas
  I.,  2011, A\&A, 536, L9

\bibitem[\protect\citeauthoryear{Tingley, Parviainen, Gandolfi, Deeg, Palle,
  {Monta\~{n}\'{e}s Rodriguez}, Murgas, Alonso, Bruntt \& Fridlund}{Tingley
  et~al.}{2014}]{Tingley2014a}
Tingley B.,  Parviainen H.,  Gandolfi D.,  Deeg H.~J.,  Palle E.,
  {Monta\~{n}\'{e}s Rodriguez} P.,  Murgas F.,  Alonso R.,  Bruntt H.,
  Fridlund M.,  2014, A\&A, 567

\end{thebibliography}

\end{document}